\documentclass[graybox]{svmult}

\usepackage{mathptmx}       
\usepackage{helvet}         
\usepackage{courier}        
\usepackage{color}
\usepackage{makeidx}         
\usepackage{graphicx}        
\usepackage{multicol}        
\usepackage[bottom]{footmisc}

\makeindex             

\begin{document}

\title*{Recent Technical Improvements to the HAYSTAC Experiment}
\author{L. Zhong, B.M. Brubaker, S.B. Cahn and S.K. Lamoreaux}
\institute{L. Zhong \at Yale University, Physics Department, 217 Prospect St, New Haven, CT 06437 U.S.A., \email{l.zhong@yale.edu}}
%
%
\maketitle

\abstract*{We report here several technical improvements to the HAYSTAC (Haloscope at Yale Sensitive To Axion Cold dark matter) that have improved operational efficiency, sensitivity, and stability.}

\abstract{We report here several technical improvements to the HAYSTAC (Haloscope at Yale Sensitive To Axion Cold dark matter) that have improved operational efficiency, sensitivity, and stability.}

\section{Overview}
\label{sec:1}
The HAYSTAC (Haloscope at Yale Sensitive To Axion Cold dark matter) was commissioned in January 2016 and has been operational for well over a year.  The experiment employs a high-$Q$ tunable microwave cavity that is immersed in a strong magnetic field (9\,T).  Putative galactic halo axions convert to radiofrequency (RF) photons in the strong magnetic field, and the cavity serves as an (imperfect) impedance matching network that couples the near infinite impedance signal source to a coaxial cable (this can be understood as an extension of the Purcell effect, as originally conceived \cite{0}), which in turn delivers the RF power to a Josephson Parametric Amplifier (JPA). Because the axion mass, hence RF frequency, is not known, the cavity and amplifier need to be tunable. The experiment thus requires a slow search over frequency for an excess RF noise signal due to axion conversion that would appear as an addition to expected quantum fluctuation noise (along with minimal thermal noise).

The first data run was completed in August 2016, with analysis results reported in \cite{1} and a detailed description of the apparatus published in \cite{2}.  In the course of this data run, several problems were identified.  First, the Kevlar pulley system that was used to vary the cavity resonance frequency by rotating the tuning rod internal to the cavity had considerable hysteresis and would drift over about 20 minutes after taking a small frequency step (typically 100\,kHz, corresponding to 0.003$^\circ$ rotation), due mainly to stiction in the tuning rod bearings. Second, the tuning rod was supported solely by thin alumina tubes that did not provide a sufficient thermal link to the tuning rod which ended up stuck at a temperature around 600\,mK, far above the base temperature of 125\,mK.  Finally, the use of thick Cu elements in the construction of the cavity support framework led to major damage of the experiment from the eddy current forces resulting from a superconducting magnet quench during a power outage in March 2016. In this note we will describe improvements to the experiment that address these problems.

\section{Attocube Motor Tuning}
\label{sec:3}
Because of the time-dependent drift after stepping the frequency using the Kevlar line pulley tuning system, we only made large frequency steps with that system, and used the insertion of a thin dielectric rod to perform the required fine stepping. Unfortunately, the range of tuning with the dielectric rod depends on frequency, and in some regions has no effect at all. Motion of the dielectric rod generated significantly more heat than the pulley system tuning.

To eliminate the stiction and hysteresis problems, we replaced the Kelvar line pulley tuning system with an Attocube ANR240 piezoelectric precision rotator~\cite{4}.  The rotator is supported by a bracket attached to the experiment frame, about 12" above the cavity.  The rotary motion is transmitted by 6" long 0.25" brass rods, connected by a corrugated stainless flexible shaft coupler.

The rotator requires a sawtooth-voltage electrical signal (with amplitude 45\,V) and draws a high current (around 1.5\,A) when actuated.
We therefore spatially separate the two wires required by the rotator from the delicate electronics wiring (flux bias current, HEMT amplifier controls, thermometry, heater and microwave switch) and provide a separate vacuum feedthrough. The resistance of the rotator wiring must be kept low, so we used 28 AWG Cu from room temperature to the 4\,K stage, and NbTi superconducting wire below the 4\,K stage.  Because the superconducting wire has too much series resistance at room temperature, the rotator cannot be tested as-wired before cooling down the system. Room temperature tests of the rotator alone are possible by use of a temporary all-Cu low resistance wire pair.

The Attocube rotator has sufficient torque to move the cavity tuning rod even in the presence of the 9\,T field.
Empirically the mechanical stiction depends on the direction of rotation. At 9\,T, we find that we are only able to tune in one direction at certain angles where the stiction is large. However, unidirectional tuning is sufficient for an axion search, and we can tune freely in both directions by reducing the magnetic field to 6\,T.

The tuning system heat load which was mainly due to large dielectric rod motions with the previous system has been improved with the current system.  It should be noted that the Kevlar line pulley system generates less heat than the Attocube system, however the latter has provided seamless operation with an acceptable heat load and no significant drift.  It is likely that even if the bearing stiction problem is solved, a drift issue would remain with the pulley system, albeit at a much reduced and possibly acceptable level.

\section{The Hot Rod Problem and Solution}
\label{sec:2}
During the commissioning phase, it was noted that the noise power at the cavity resonant frequency was always significantly higher than the noise power observed off-resonance. The off-resonance noise was consistent with the expected system noise temperature.  Various tests, including raising the system temperature to the point where the cavity and off-resonance noise levels were almost equal, provided firm evidence that the excess noise was due to the tuning rod (or other system element) being at a high temperature.  These tests alleviated concerns that the excess noise was due to a spurious interaction between the cavity and JPA which might have been difficult to eliminate. Such a large-volume and uniquely configured cavity had never before been coupled to a JPA.

We had long known that the thermal link to the cavity tuning rod was at best weak, being several inches of  0.250" OD, 0.125" ID polycrystalline alumina tubes, at least an inch long, on either end of the cavity cylinder axis.  The only contact between the tubes and the support frame (which serves as the thermal link to the dilution refrigerator mixing chamber) were precision ball bearings that were used to ensure as free a rotation of the rod as practical.  The contact area between the balls and the races is vanishingly small by design, further complicating the thermal link problem.  We had recognized this problem earlier, and attempted to provide a link by gluing (using thermal epoxy) a short brass rods (0.25") into the external ends of the alumina tubes, and then connecting those brass rods to the support frame with flexible Cu braids.

Tests at U.C. Berkeley suggested that a 0.125" Cu rod could be inserted far into alumina tubes (to a depth where they are within the cavity), without undue loss of cavity $Q$~\cite{3}. Such rods were incorporated, inserted as far as possible into the alumina tubes, and glued in place using conductive silver epoxy (Epo-Tek H20E). The rods are long enough so a 1/2" nub extends beyond the tube, and Cu braids were soldered onto these nubs before gluing. The upper nub serves as the connection point to the piezoelectric rotator (discussed above).

According to the latest measurements, the Cu rods reduce the total system noise photon number at the cavity resonant frequency from around 3 to 2.3 on average, corresponding to a reduction in tuning rod temperature from 600\,mK to 250\,mK. Unfortunately, the cavity $Q$ has been reduced by about 40\%.

Incorporation of the Cu rod thermal links reduced the time to cool the system from when the mixture is first condensed to the base temperature from over six hours to under one hour.  This was because the alumina tubes' weak thermal link became weaker with reduction in temperature, becoming a bottleneck in the cooling process while maintaining a substantial heat load. The alumina tubes became effective thermal insulators when the end affixed to the tuning rod reached 600\,mK (with the other end at the 125\,mK frame temperature). After this quasi-equilibrium was established, we saw no discernible decrease in thermal noise level over months of operation, implying a time of perhaps years for the tuning rod to significantly cool beyond this point. We have yet to identify the remaining source of excess thermal noise (250\,mK compared to a system temperature of 125\,mK) which is likely a further issue with the tuning rod thermal link.

An unfortunate consequence of the thermal links is that noise is coupled into the cavity.  The links act as antennas that couple signals directly into the cavity, as might be expected because the mechanisms that result in reduction in $Q$ likely have a component due to the internal field leaking, and radiating, to free space. Although one might expect the radiofrequency (RF) noise within the cryostat to be extremely low, there apparently is enough noise (mostly from coherent sources, so technically it is not noise but spurious or systematic signals) to be problematic. There are many possible noise sources, including coupling between the room temperature part of the cryostat and the cavity region due to signals running on thermometer wires, the outsides of coaxial cables, or the cryostat support structure and piping.  RF noise can also couple into the cryostat through the myriad thermometer, heater, and control wires.  Although there are filters in place that offer more than 100 dB of attenuation, we must recall that the system is sensitive at the single photon or Yoctowatt level. The current operating frequency of 5.6-5.7 GHz is the so-called 5 GHz WiFi band, and our impression is that this is the source of most of the noise.

The RF noise problem can be solved by enclosing the thermal links in a cylindrical shield that attaches to the cavity and surrounds the rods, with the flexible braid attached to the inner surface of the shield, and an endcap on the cylinder.

\section{Cu Plated Stainless Thermal Links and Shields}
\label{sec:4}

Our original design incorporated many massive OFCH Cu components, and as mentioned in the introduction, the forces that were generated during a magnet quench led to significant damage to the experiment.  We have noticed that the cavity, fabricated from stainless steel and Cu plated, apparently has a very high thermal conductivity.  Thermometers are mounted on the cavity top (where the thermal link to the frame is located) and bottom, and practically there is almost no time delay (less than 30 seconds) between a temperature change at the top, and subsequent change at the bottom.

The damaged still-temperature thermal shield was replaced with a Cu plated (to 0.002") stainless steel shield \cite{4a}. This new shield has been sufficient with no obvious excess heat load at the mixing chamber level.  We have yet to see its response to a magnet quench, however we plan to replace all massive Cu parts with Cu plated stainless steel in the next upgrade to the experiment.

\section{Conclusion}
\label{sec:4}
Incorporation of the Attocube rotator has solved the hysteresis and drift problem of previous Kevlar pulley system, and reduced the complexity of operation.  The addition of the Cu rods has provided better cooling efficiency of the cavity tuning rod at the cost of 40\% reduction of cavity $Q$: the net result is that operation of the system is more reliable and the scan rate is unchanged, however the rods allow RF noise to be coupled into the cavity so there are additional peaks that require extra and time-consuming attention; this problem does have a solution.  In the near future, we plan to upgrade the experiment, which will include moving it to a new BlueFors dilution refrigerator, the replacement of massive Cu components in the high magnetic field regions with Cu plated stainless steel, and the incorporation of a squeezed state receiver system as describe in \cite{5}.

\section{Acknowledgments}
This  work  was  supported  by  the  National  Science Foundation, under grants PHY-1362305 and PHY-1607417, by the Heising-Simons Foundation under grants 2014-181, 2014-182, and 2014-183, and by Yale University.  HAYSTAC is a collaboration between U.C. Berkeley, Colorado University/JILA, and Yale.

\end{document}